\documentclass[draft]{agujournal2019}
\usepackage{subfigure}
\usepackage[english]{babel}
\usepackage{ragged2e}
\usepackage[none]{hyphenat}
\draftfalse

\journalname{Seismological Research Letters}

\begin{document}
%
%

\title{Imaging urban hidden faults with ambient noise recorded by dense seismic arrays}

%
%




\authors{E. Biondi\affil{1}\thanks{1200 E California Blvd, Pasadena, CA 91125}, Jorge C. Castellanos\affil{2}, Robert W. Clayton\affil{1}}


\affiliation{1}{California Institute of Technology, Seismological Laboratory; 1200 E. California Blvd., MS 252-21 Pasadena, California 91125-2100}

\affiliation{2}{Risk Management Solutions; 7575 Gateway Blvd., Suite 300, Newark, CA 94560, USA}




\correspondingauthor{Ettore Biondi}{ebiondi@caltech.edu}





%
%

%
%


\begin{abstract}
\justifying
The identification of preexisting near-surface faults represents a piece of crucial information needed to correctly assess the seismic hazard of any area. The mapping of these structures is particularly challenging in densely populated and heavily urbanized areas. We use ambient seismic noise recorded by a dense array in Seal Beach, California, to image shallow fault lines via a reflected surface wave analysis. Our results highlight the presence of previously unknown shallow faults that correlate remarkably well with shallow seismicity and active survey images.
\end{abstract}

\section*{Introduction}

The seismic hazard in earthquake-prone areas such as Southern California is highly dependent on the near-surface properties and pre-existing geologic structures~\cite{wald1998seismic,beroza1991near,graves2008seismic,jin2021importance,castellanos2021fine}. The analysis of earthquake catalogs provides invaluable information on the presence of active fault lines~\cite{storchak2013public,ross2019searching}, while subsurface imaging and velocity estimation procedures generate the necessary material properties to estimate ground shaking models~\cite{frankel1999does,olsen2006strong,graves2016kinematic}. Therefore, the accurate characterization of the near subsurface is crucial to better assess earthquake-induced ground shaking, which ultimately would lead to a more resilient urban infrastructure~\cite{olsen2009shakeout}. Even though there has been great progress in the techniques and instrumentations used for the characterization of the near-fault and near-surface structures~\cite{socco2004surface,atterholt2022fault,liu2021ambient,yao2011structure,fichtner2019theoretical}, it is still challenging to properly identify existing geological features posing risks in densely populated areas. 

Los Angeles area in California represents one of the most studied zones affected by earthquakes~\cite{harris1995influence,allen2003potential,hauksson2012waveform}. In this area, the main seismic hazard is controlled by the Newport-Inglewood Fault (NIF) zone~\cite{field2015long,taber1920inglewood}, which is believed to have ruptured during the damaging magnitude 6.4 1933 Long Beach earthquake~\cite{wood1933long,hauksson1991source,hough20201933}. The NIF is predominantly characterized as a strike-slip fault, having an accumulated displacement of approximately 60 kilometers~\cite{hauksson1987seismotectonics}. However, the low level of seismic activity associated with this fault zone complicates the precise delineation of its structural contours. Moreover, most of its surface expressions are effectively being erased by the rapid urban development of the area~\cite{shaw1996earthquake}. Several studies employed data from temporary dense seismic arrays to characterize the subsurface structures of this area~\cite{shaw1996earthquake,lin2013high,nakata2015body,castellanos2020using,jia2021determination}. Although these analyses provided important information on the subsurface wave speeds, they were not able to constrain the near-surface fault structures at high resolution. This limitation is due to the resolution commonly achieved by surface-wave inversion algorithms preventing the identification of any sharp change in material properties. By employing scattered waves, our method overcomes this resolution barrier and can image such contrasts in a great level of detail.

In this study, we employ data from one of these temporary arrays deployed in the Los Angeles area to image the near-surface fault structures with a new level of detail compared to previous studies. Using the passive seismic noise recorded by the considered nodal dense array, we compute cross-correlation functions that reconstruct the propagation of surface waves across the array~\cite{curtis2006seismic}. The correlograms we obtained reveal scattered waves that yield crucial insights into subsurface features, particularly faults. By leveraging the dense arrangement of the seismic array, we apply a novel imaging technique to localize these formations. Our approach facilitates the mapping of existing faults, which in turn permits a better seismic hazard assessment in heavily urbanized zones.

\section*{Methods}

\subsection*{Precursory energy imaging procedure}
We show how our method performs on synthetic data in which a known seismic-energy scattering structure is included within a surface velocity model. To simulate noise cross-correlations, we compute acoustic isotropic Green’s functions using a finite-difference approach from sources recorded on a line of stations (Figure 1(a))~\cite{barnier2023full}. The source impulse response is a Ricker wavelet with a central frequency of 0.5 Hz. The velocity model considered has a constant velocity of 2 km/s with a linear high-velocity feature representing a fault line (Figure 1(b)). For each source, we compute the cross-correlations for all station pairs and stacked them using a non-uniform source power to mimic the ocean microseism. For a virtual source at the edge of the station line, the cross-correlation clearly shows the direct arrival, the direct scattering, and the precursory arrival considered in our method for imaging purposes (Figure 1(c)). When a zero-lag cross-section of the correlograms for all virtual source-receiver pairs is visualized, we observe a clear separation between the direct and precursor arrivals (Figure 1(d)). This separation allows us to easily mute the direct arrival and sum the envelope of the precursory energy along the direction orthogonal with respect to the zero-offset line (i.e., source and receiver placed at the same location). This process coherently stacks any energy generated by scattered waves and forms a peak at the location of the scattering structure (Figure 1(e)). When applied to a dense array of seismic stations, such as the one used in this study, along multiple lines, we can form a 2D image of the near-surface scattering structures (e.g., fault lines).

\subsection*{Data preprocessing and cross-correlation computation}
To obtain the necessary station-pair cross-correlations for our imaging procedure, we first apply a spectral whitening operation and compute the correlation within the frequency domain~\cite{groos2012performance}. This process is performed on hourly segments of continuous data and stacked until convergence to a stable correlation function is reached. We can obtain a stable and acceptable signal-noise level in the correlograms using one week of data but for the reported study we employ the full month available. The nodal data are composed of single vertical component instruments; thus, the obtained correlations are mostly due to the recorded Rayleigh surface waves. Since our method works on lines of stations, we compute and process cross-correlations using stripes of stations with a width of 400 m, while the binning process along the line direction is performed using a width of 150 m.

\section*{Results}
\subsection*{Cross-correlation precursor fault imaging}
The seismic data employed in our study were recorded by a dense array of Fairfield Z‐land nodes each consisting of a vertical short‐period velocity sensor with a lower‐frequency corner of 10 Hz deployed in the Seal Beach area of Los Angeles, California (Figure 2). The dataset is composed of one month of data recorded in early 2018 and a previous study has used these recordings to determine for the subsurface velocity structure using noise interferometry~\cite{castellanos2021fine}. This area presents a complex fault system that has been recently interpreted by a private geophysical company for exploration purposes~\cite{yang2023shallow,gish2023sealbeach} (dash lines in Figure 2). Additionally, two major fault lines cross this area that are mapped by the United States Geological Survey (USGS)~\cite{frankel2000usgs}: the Los Alamitos (LAF) and the Newport‐Inglewood (NIF) faults. The continuous recordings from this nodal array are used to compute noise cross-correlation functions that include spectral whitening~\cite{shapiro2004emergence,groos2012performance} (Figure 3(a)). To better identify the presence of any precursory signals within the noise cross-correlation functions, we compute the correlations along a narrow profile (Figure 2(b)). By binning the stations along this line, we can create a 3D volume as a function of time lag, receiver position, and virtual source location along this line, which can be effectively considered as an active seismic survey~\cite{claerbout2008basic}. The direct arrival is aligned along a 45º angle direction for constant time-lag slices of the volume (Figure 3(c), Supplementary movies 1 and 2), while, at zero-time lag, any precursory signals would appear away from this diagonal and approximately orthogonally to the direct wave. Our stacking procedure allows us to identify any structure that scatters surface waves located in the proximity of the imaging line (Figure 3(d)). Therefore, any scattering structure crossing the considered line, such as faults, can be imaged via our noise-based procedure. In particular, we assume that scatterers are composed of line segments whose scattering potential is maximum for waves impinging orthogonally to them~\cite{yi2023thesis}.   

\subsection*{The shallow faults in the Seal Beach area}
The application of our method to multiple parallel lines of nodes allows us to reveal the presence of shallow faults beneath the Seal Beach array (Figure 4). The panel of Figure 4(a) shows the image obtained using the lowest frequency band considered (0.5-1.0 Hz), where the surface waves scattered by subsurface structures related to the Newport-Inglewood fault are imaged in the north-west portion of the array. In a higher–frequency band, images clearly depict the presence of shallower faults in the area. In particular, the unmapped fault parallel with the previously known major structures is visible in most panels (Figures 4(b)-(e)), highlighting its presence in the near-surface. Moreover, the locations of the shallow earthquakes align remarkably well with our newly imaged structure. This observation also applies to the seismicity and structures imaged in the near-shore area in the southeast portion of the array (Figures 4(d)-(e)) ~\cite{yang2023shallow}. These structures are likely shallower than the mapped faults since their expression is only visible in the highest-frequency band images.

In addition to the correspondence of our images with the shallow seismicity, the presence of the unmapped shallow fault of Figure 4(b) is also verified by comparing an image line crossing this structure with a profile of an active-source seismic study performed in this area (Figure 5) ~\cite{wilson2015characterization}. At approximately 5.9 km distance along the profile (red dashed line in Figure 5(a)), the stacked scattered energy shows a peak (Figure 5(b)) in correspondence with the shallow fault interpreted on the seismic image shown in Figure 5(c). The deeper Garden Grove fault (centered at approximately 4 km distance on the profile) is depicted by the broad peak in stacked energy for the lowest frequency band considered in our imaging process. It is also noticeable the remarkable correlation between the complex structures present southwest of the Garden Grove fault and the increase in the number of peaks, which highlights the ability of our method to detect portions of the near-surface heavily damaged. Similar correlations are observable for other seismic profiles (Supplementary Figures 1 and 2). This observation is particularly relevant for earthquake hazard assessments. Some damaged fault zones, for instance, have been observed to act as barriers to rupture propagation~\cite{aki1979characterization}. They are also often sites of aftershocks, which makes them important for understanding such earthquake sequences. Finally, the NIF is part of the Alquist-Priolo fault zones and it is assumed to be a few hundred meters wide~\cite{reitherman1992effectiveness}; however, the profile shown in Figure 5 highlights the presence of a significantly wider area stretching over a few kilometers of the Seal Beach area. The presence of this damaged zone as well as the near-surface faults change the susceptibility of this area to shaking in case of a major seismic event due to their ability to trap wave energy~\cite{li1990fault}. 

Our images correlate well with the previously interpreted major fault structures in the area~\cite{hauksson1990earthquakes}. The Newport-Inglewood fault is predominantly a strike-slip system with portions affected by compression causing this system to become a complex flower structure in which multiple secondary strands depart from the deeper main fault~\cite{yeats1973newport}. These secondary fault branches are shallower and thus act as surface-wave scatterers that are captured by our energy-based images (Figure 5). On the other hand, the blind Los Alamitos fault in the east portion of the dense array (Figure 2) does not show a clear expression in our images. This fault represents the bounding structure confining the deep Los Angeles sedimentary basin on the east of this area and it is interpreted to be deeper than the Newport-Inglewood ruptures~\cite{wilson2015characterization}. Therefore, the considered frequency bands in this study may not have the necessary sensitivity to form an image of this structure. 

\section*{Discussion}

The depth sensitivity of surface waves depends on their frequency content~\cite{herrmann2013computer}. This behavior allows us to infer the depth range of the features imaged in our results. To quantify this effect, we perform a synthetic experiment using an elastic isotropic finite-different simulation~\cite{biondi2021target}. We apply a free-surface boundary condition on the top part of the model and absorbing layers in the other sections of the domain~\cite{robertsson1996numerical} and employ an explosive source placed at 10 m depth with a frequency content between 0.5 to 3.5 Hz. The velocity model considered is a 1D profile in which five high-velocity anomalies representing scattering features are placed at different depths (Figure 6(a)). Surface waves at lower frequencies are affected by deeper structures (Figure 6(b)), while frequencies higher than 2 Hz can provide information about the first 100 m in the subsurface (Figures 6(c)-(d)). This effect is also clear by looking at the seismograms at different frequencies recorded by the surface array (Supplementary Figure 3). Therefore, the sensitivity kernels of Figure 6(e) can constrain the maximum depth of the unmapped fault to approximately 200 m depth (Figure 4(b)), which is also verified by the active survey image (Figure 5(c)). On the other hand, the faults present near the shoreline are likely to be confined within the first 100 m of the near subsurface.

The ability of our imaging method depends on the binning direction of the stations considered within the cross-correlation procedure. To highlight the effect of how this parameter affects our results, we form an image using station lines in the east-west direction (Supplementary Figure 4(a)). In this image, we cannot observe the energy from the unmapped faults, which is due to the stacking direction within the zero-time-lag panel. Conversely, parts of other near-surface faults are visible in the west portion of the Seal Beach array. This test demonstrates that the optimal direction to form images of fault with our approach is the one orthogonal to the supposed structure.
Seismic sources constantly generating noise could also give rise to precursory signals in cross-correlations~\cite{ma2013locating,retailleau2017locating}. For this reason, we also verify that our precursory energy that is employed for image formation is generated by scattered surface waves and not secondary urban noise sources~\cite{guearthquake}. The lowest noise level observed in the Seal Beach array is within the night hours when the lowest car movements are present and when the major source of noise is the ocean microseism~\cite{ardhuin2015ocean}. The imaging procedure using only the night hours generates almost identical images to ones using the entire day (Supplementary Figures 4(b) and 4(c)), giving us confidence that the structures appearing within our plots represent subsurface features scattering Rayleigh surface waves.

Our approach not only underscores the adaptation of ambient noise for seismic exploration but also marks a leap forward in urban seismic hazard assessment, exemplifying the potential of turning the ordinary hum of the environment into a powerful tool for unveiling the hidden structures of the near-surface Earth's crust.

\section*{Conclusions}
We showcase the utilization of scattered energy embedded within seismic noise cross-correlations to construct images of subsurface structures. By considering multiple linear arrays of stations, we can clearly identify any scattered waves that using a simple beamforming mechanism permit the localization of the causing structures. When our method is applied to a dense seismic array in Seal Beach, we can delineate damage zones linked to the NIF and identify an unmapped shallow fault structure The interpretation of such structures is also corroborated by the previously observed shallow seismicity and the images obtained by an active seismic survey.

\bibliography{main}

\begin{thebibliography}{}

\bibitem [\protect \citeauthoryear {%
Aki%
}{%
Aki%
}{%
{\protect \APACyear {1979}}%
}]{%
aki1979characterization}
\APACinsertmetastar {%
aki1979characterization}%
\begin{APACrefauthors}%
Aki, K.%
\end{APACrefauthors}%
\unskip\
\newblock
\APACrefYearMonthDay{1979}{}{}.
\newblock
{\BBOQ}\APACrefatitle {Characterization of barriers on an earthquake fault} {Characterization of barriers on an earthquake fault}.{\BBCQ}
\newblock
\APACjournalVolNumPages{Journal of Geophysical Research: Solid Earth}{84}{B11}{6140--6148}.
\PrintBackRefs{\CurrentBib}

\bibitem [\protect \citeauthoryear {%
Allen%
\ \BBA {} Kanamori%
}{%
Allen%
\ \BBA {} Kanamori%
}{%
{\protect \APACyear {2003}}%
}]{%
allen2003potential}
\APACinsertmetastar {%
allen2003potential}%
\begin{APACrefauthors}%
Allen, R\BPBI M.%
\BCBT {}\ \BBA {} Kanamori, H.%
\end{APACrefauthors}%
\unskip\
\newblock
\APACrefYearMonthDay{2003}{}{}.
\newblock
{\BBOQ}\APACrefatitle {The potential for earthquake early warning in southern {C}alifornia} {The potential for earthquake early warning in southern {C}alifornia}.{\BBCQ}
\newblock
\APACjournalVolNumPages{Science}{300}{5620}{786--789}.
\PrintBackRefs{\CurrentBib}

\bibitem [\protect \citeauthoryear {%
Ardhuin%
, Gualtieri%
\BCBL {}\ \BBA {} Stutzmann%
}{%
Ardhuin%
\ \protect \BOthers {.}}{%
{\protect \APACyear {2015}}%
}]{%
ardhuin2015ocean}
\APACinsertmetastar {%
ardhuin2015ocean}%
\begin{APACrefauthors}%
Ardhuin, F.%
, Gualtieri, L.%
\BCBL {}\ \BBA {} Stutzmann, E.%
\end{APACrefauthors}%
\unskip\
\newblock
\APACrefYearMonthDay{2015}{}{}.
\newblock
{\BBOQ}\APACrefatitle {{How ocean waves rock the Earth: Two mechanisms explain microseisms with periods 3 to 300 s}} {{How ocean waves rock the Earth: Two mechanisms explain microseisms with periods 3 to 300 s}}.{\BBCQ}
\newblock
\APACjournalVolNumPages{Geophysical Research Letters}{42}{3}{765--772}.
\PrintBackRefs{\CurrentBib}

\bibitem [\protect \citeauthoryear {%
Atterholt%
, Zhan%
\BCBL {}\ \BBA {} Yang%
}{%
Atterholt%
\ \protect \BOthers {.}}{%
{\protect \APACyear {2022}}%
}]{%
atterholt2022fault}
\APACinsertmetastar {%
atterholt2022fault}%
\begin{APACrefauthors}%
Atterholt, J.%
, Zhan, Z.%
\BCBL {}\ \BBA {} Yang, Y.%
\end{APACrefauthors}%
\unskip\
\newblock
\APACrefYearMonthDay{2022}{}{}.
\newblock
{\BBOQ}\APACrefatitle {{Fault Zone Imaging With Distributed Acoustic Sensing: Body-To-Surface Wave Scattering}} {{Fault Zone Imaging With Distributed Acoustic Sensing: Body-To-Surface Wave Scattering}}.{\BBCQ}
\newblock
\APACjournalVolNumPages{Journal of Geophysical Research: Solid Earth}{127}{11}{e2022JB025052}.
\PrintBackRefs{\CurrentBib}

\bibitem [\protect \citeauthoryear {%
Barnier%
, Biondi%
, Clapp%
\BCBL {}\ \BBA {} Biondi%
}{%
Barnier%
\ \protect \BOthers {.}}{%
{\protect \APACyear {2023}}%
}]{%
barnier2023full}
\APACinsertmetastar {%
barnier2023full}%
\begin{APACrefauthors}%
Barnier, G.%
, Biondi, E.%
, Clapp, R\BPBI G.%
\BCBL {}\ \BBA {} Biondi, B.%
\end{APACrefauthors}%
\unskip\
\newblock
\APACrefYearMonthDay{2023}{}{}.
\newblock
{\BBOQ}\APACrefatitle {Full waveform inversion by model extension: practical applications} {Full waveform inversion by model extension: practical applications}.{\BBCQ}
\newblock
\APACjournalVolNumPages{Geophysics}{88}{5}{1--138}.
\PrintBackRefs{\CurrentBib}

\bibitem [\protect \citeauthoryear {%
Beroza%
}{%
Beroza%
}{%
{\protect \APACyear {1991}}%
}]{%
beroza1991near}
\APACinsertmetastar {%
beroza1991near}%
\begin{APACrefauthors}%
Beroza, G\BPBI C.%
\end{APACrefauthors}%
\unskip\
\newblock
\APACrefYearMonthDay{1991}{}{}.
\newblock
{\BBOQ}\APACrefatitle {Near-source modeling of the {L}oma {P}rieta earthquake: Evidence for heterogeneous slip and implications for earthquake hazard} {Near-source modeling of the {L}oma {P}rieta earthquake: Evidence for heterogeneous slip and implications for earthquake hazard}.{\BBCQ}
\newblock
\APACjournalVolNumPages{Bulletin of the Seismological Society of America}{81}{5}{1603--1621}.
\PrintBackRefs{\CurrentBib}

\bibitem [\protect \citeauthoryear {%
Biondi%
}{%
Biondi%
}{%
{\protect \APACyear {2021}}%
}]{%
biondi2021target}
\APACinsertmetastar {%
biondi2021target}%
\begin{APACrefauthors}%
Biondi, E.%
\end{APACrefauthors}%
\unskip\
\newblock
\APACrefYear{2021}.
\newblock
\APACrefbtitle {Target-oriented elastic full-waveform inversion} {Target-oriented elastic full-waveform inversion}.
\newblock
\APACaddressPublisher{}{Stanford University}.
\PrintBackRefs{\CurrentBib}

\bibitem [\protect \citeauthoryear {%
Castellanos%
\ \BBA {} Clayton%
}{%
Castellanos%
\ \BBA {} Clayton%
}{%
{\protect \APACyear {2021}}%
}]{%
castellanos2021fine}
\APACinsertmetastar {%
castellanos2021fine}%
\begin{APACrefauthors}%
Castellanos, J\BPBI C.%
\BCBT {}\ \BBA {} Clayton, R\BPBI W.%
\end{APACrefauthors}%
\unskip\
\newblock
\APACrefYearMonthDay{2021}{}{}.
\newblock
{\BBOQ}\APACrefatitle {The fine-scale structure of {L}ong {B}each, {C}alifornia, and its impact on ground motion acceleration} {The fine-scale structure of {L}ong {B}each, {C}alifornia, and its impact on ground motion acceleration}.{\BBCQ}
\newblock
\APACjournalVolNumPages{Journal of Geophysical Research: Solid Earth}{126}{12}{e2021JB022462}.
\PrintBackRefs{\CurrentBib}

\bibitem [\protect \citeauthoryear {%
Castellanos%
, Clayton%
\BCBL {}\ \BBA {} Juarez%
}{%
Castellanos%
\ \protect \BOthers {.}}{%
{\protect \APACyear {2020}}%
}]{%
castellanos2020using}
\APACinsertmetastar {%
castellanos2020using}%
\begin{APACrefauthors}%
Castellanos, J\BPBI C.%
, Clayton, R\BPBI W.%
\BCBL {}\ \BBA {} Juarez, A.%
\end{APACrefauthors}%
\unskip\
\newblock
\APACrefYearMonthDay{2020}{}{}.
\newblock
{\BBOQ}\APACrefatitle {{Using a time-based subarray method to extract and invert noise-derived body waves at Long Beach, California}} {{Using a time-based subarray method to extract and invert noise-derived body waves at Long Beach, California}}.{\BBCQ}
\newblock
\APACjournalVolNumPages{Journal of Geophysical Research: Solid Earth}{125}{5}{e2019JB018855}.
\PrintBackRefs{\CurrentBib}

\bibitem [\protect \citeauthoryear {%
Claerbout%
\ \BBA {} Green%
}{%
Claerbout%
\ \BBA {} Green%
}{%
{\protect \APACyear {2008}}%
}]{%
claerbout2008basic}
\APACinsertmetastar {%
claerbout2008basic}%
\begin{APACrefauthors}%
Claerbout, J\BPBI F.%
\BCBT {}\ \BBA {} Green, I.%
\end{APACrefauthors}%
\unskip\
\newblock
\APACrefYearMonthDay{2008}{}{}.
\newblock
\APACrefbtitle {Basic earth imaging.} {Basic earth imaging.}
\newblock
\APACaddressPublisher{}{Citeseer}.
\PrintBackRefs{\CurrentBib}

\bibitem [\protect \citeauthoryear {%
Curtis%
, Gerstoft%
, Sato%
, Snieder%
\BCBL {}\ \BBA {} Wapenaar%
}{%
Curtis%
\ \protect \BOthers {.}}{%
{\protect \APACyear {2006}}%
}]{%
curtis2006seismic}
\APACinsertmetastar {%
curtis2006seismic}%
\begin{APACrefauthors}%
Curtis, A.%
, Gerstoft, P.%
, Sato, H.%
, Snieder, R.%
\BCBL {}\ \BBA {} Wapenaar, K.%
\end{APACrefauthors}%
\unskip\
\newblock
\APACrefYearMonthDay{2006}{}{}.
\newblock
{\BBOQ}\APACrefatitle {{Seismic interferometry—Turning noise into signal}} {{Seismic interferometry—Turning noise into signal}}.{\BBCQ}
\newblock
\APACjournalVolNumPages{The Leading Edge}{25}{9}{1082--1092}.
\PrintBackRefs{\CurrentBib}

\bibitem [\protect \citeauthoryear {%
Fichtner%
, Tsai%
, Nakata%
\BCBL {}\ \BBA {} Gualtieri%
}{%
Fichtner%
\ \protect \BOthers {.}}{%
{\protect \APACyear {2019}}%
}]{%
fichtner2019theoretical}
\APACinsertmetastar {%
fichtner2019theoretical}%
\begin{APACrefauthors}%
Fichtner, A.%
, Tsai, V.%
, Nakata, N.%
\BCBL {}\ \BBA {} Gualtieri, L.%
\end{APACrefauthors}%
\unskip\
\newblock
\APACrefYearMonthDay{2019}{}{}.
\newblock
{\BBOQ}\APACrefatitle {Theoretical foundations of noise interferometry} {Theoretical foundations of noise interferometry}.{\BBCQ}
\newblock
\APACjournalVolNumPages{Seismic ambient noise}{}{}{109--143}.
\PrintBackRefs{\CurrentBib}

\bibitem [\protect \citeauthoryear {%
Field%
\ \protect \BOthers {.}}{%
Field%
\ \protect \BOthers {.}}{%
{\protect \APACyear {2015}}%
}]{%
field2015long}
\APACinsertmetastar {%
field2015long}%
\begin{APACrefauthors}%
Field, E\BPBI H.%
, Biasi, G\BPBI P.%
, Bird, P.%
, Dawson, T\BPBI E.%
, Felzer, K\BPBI R.%
, Jackson, D\BPBI D.%
\BDBL {}others%
\end{APACrefauthors}%
\unskip\
\newblock
\APACrefYearMonthDay{2015}{}{}.
\newblock
{\BBOQ}\APACrefatitle {{Long-term time-dependent probabilities for the third Uniform California Earthquake Rupture Forecast (UCERF3)}} {{Long-term time-dependent probabilities for the third Uniform California Earthquake Rupture Forecast (UCERF3)}}.{\BBCQ}
\newblock
\APACjournalVolNumPages{Bulletin of the Seismological Society of America}{105}{2A}{511--543}.
\PrintBackRefs{\CurrentBib}

\bibitem [\protect \citeauthoryear {%
A.~Frankel%
\ \protect \BOthers {.}}{%
A.~Frankel%
\ \protect \BOthers {.}}{%
{\protect \APACyear {2000}}%
}]{%
frankel2000usgs}
\APACinsertmetastar {%
frankel2000usgs}%
\begin{APACrefauthors}%
Frankel, A.%
, Mueller, C.%
, Barnhard, T.%
, Leyendecker, E.%
, Wesson, R.%
, Harmsen, S.%
\BDBL {}others%
\end{APACrefauthors}%
\unskip\
\newblock
\APACrefYearMonthDay{2000}{}{}.
\newblock
{\BBOQ}\APACrefatitle {{USGS national seismic hazard maps}} {{USGS national seismic hazard maps}}.{\BBCQ}
\newblock
\APACjournalVolNumPages{Earthquake spectra}{16}{1}{1--19}.
\PrintBackRefs{\CurrentBib}

\bibitem [\protect \citeauthoryear {%
A\BPBI D.~Frankel%
}{%
A\BPBI D.~Frankel%
}{%
{\protect \APACyear {1999}}%
}]{%
frankel1999does}
\APACinsertmetastar {%
frankel1999does}%
\begin{APACrefauthors}%
Frankel, A\BPBI D.%
\end{APACrefauthors}%
\unskip\
\newblock
\APACrefYearMonthDay{1999}{}{}.
\newblock
{\BBOQ}\APACrefatitle {How does the ground shake?} {How does the ground shake?}{\BBCQ}
\newblock
\APACjournalVolNumPages{Science}{283}{5410}{2032--2033}.
\PrintBackRefs{\CurrentBib}

\bibitem [\protect \citeauthoryear {%
Gish%
\ \BBA {} Boljen%
}{%
Gish%
\ \BBA {} Boljen%
}{%
{\protect \APACyear {2023}}%
}]{%
gish2023sealbeach}
\APACinsertmetastar {%
gish2023sealbeach}%
\begin{APACrefauthors}%
Gish, D.%
\BCBT {}\ \BBA {} Boljen, S.%
\end{APACrefauthors}%
\unskip\
\newblock
\APACrefYearMonthDay{2023}{}{}.
\newblock
{\BBOQ}\APACrefatitle {{Long Beach-Seal Beach Thrust Belt: Tectonic Evaluation of the Newport-Inglewood Fault Zone and Seal Beach Anticline Using Modern 3D Seismic}} {{Long Beach-Seal Beach Thrust Belt: Tectonic Evaluation of the Newport-Inglewood Fault Zone and Seal Beach Anticline Using Modern 3D Seismic}}.{\BBCQ}
\newblock
\APACjournalVolNumPages{Association of Petroleum Geologists Bulletin}{}{}{}.
\newblock
\APACrefnote{under revision}
\PrintBackRefs{\CurrentBib}

\bibitem [\protect \citeauthoryear {%
R.~Graves%
\ \BBA {} Pitarka%
}{%
R.~Graves%
\ \BBA {} Pitarka%
}{%
{\protect \APACyear {2016}}%
}]{%
graves2016kinematic}
\APACinsertmetastar {%
graves2016kinematic}%
\begin{APACrefauthors}%
Graves, R.%
\BCBT {}\ \BBA {} Pitarka, A.%
\end{APACrefauthors}%
\unskip\
\newblock
\APACrefYearMonthDay{2016}{}{}.
\newblock
{\BBOQ}\APACrefatitle {Kinematic ground-motion simulations on rough faults including effects of 3{D} stochastic velocity perturbations} {Kinematic ground-motion simulations on rough faults including effects of 3{D} stochastic velocity perturbations}.{\BBCQ}
\newblock
\APACjournalVolNumPages{Bulletin of the Seismological Society of America}{106}{5}{2136--2153}.
\PrintBackRefs{\CurrentBib}

\bibitem [\protect \citeauthoryear {%
R\BPBI W.~Graves%
}{%
R\BPBI W.~Graves%
}{%
{\protect \APACyear {2008}}%
}]{%
graves2008seismic}
\APACinsertmetastar {%
graves2008seismic}%
\begin{APACrefauthors}%
Graves, R\BPBI W.%
\end{APACrefauthors}%
\unskip\
\newblock
\APACrefYearMonthDay{2008}{}{}.
\newblock
{\BBOQ}\APACrefatitle {The seismic response of the {S}an {B}ernardino basin region during the 2001 {B}ig {B}ear {L}ake earthquake} {The seismic response of the {S}an {B}ernardino basin region during the 2001 {B}ig {B}ear {L}ake earthquake}.{\BBCQ}
\newblock
\APACjournalVolNumPages{Bulletin of the Seismological Society of America}{98}{1}{241--252}.
\PrintBackRefs{\CurrentBib}

\bibitem [\protect \citeauthoryear {%
Groos%
, Bussat%
\BCBL {}\ \BBA {} Ritter%
}{%
Groos%
\ \protect \BOthers {.}}{%
{\protect \APACyear {2012}}%
}]{%
groos2012performance}
\APACinsertmetastar {%
groos2012performance}%
\begin{APACrefauthors}%
Groos, J.%
, Bussat, S.%
\BCBL {}\ \BBA {} Ritter, J.%
\end{APACrefauthors}%
\unskip\
\newblock
\APACrefYearMonthDay{2012}{}{}.
\newblock
{\BBOQ}\APACrefatitle {Performance of different processing schemes in seismic noise cross-correlations} {Performance of different processing schemes in seismic noise cross-correlations}.{\BBCQ}
\newblock
\APACjournalVolNumPages{Geophysical Journal International}{188}{2}{498--512}.
\PrintBackRefs{\CurrentBib}

\bibitem [\protect \citeauthoryear {%
Gu%
, Zhang%
, Nakata%
\BCBL {}\ \BBA {} Gao%
}{%
Gu%
\ \protect \BOthers {.}}{%
{\protect \APACyear {2021}}%
}]{%
guearthquake}
\APACinsertmetastar {%
guearthquake}%
\begin{APACrefauthors}%
Gu, N.%
, Zhang, H.%
, Nakata, N.%
\BCBL {}\ \BBA {} Gao, J.%
\end{APACrefauthors}%
\unskip\
\newblock
\APACrefYearMonthDay{2021}{}{}.
\newblock
{\BBOQ}\APACrefatitle {Faultdetectionbyreflectedsurfacewavesbasedonambient noiseinterferometry} {Faultdetectionbyreflectedsurfacewavesbasedonambient noiseinterferometry}.{\BBCQ}
\newblock
\APACjournalVolNumPages{Earthquake Research Advances}{1}{4}{}.
\PrintBackRefs{\CurrentBib}

\bibitem [\protect \citeauthoryear {%
Harris%
, Simpson%
\BCBL {}\ \BBA {} Reasenberg%
}{%
Harris%
\ \protect \BOthers {.}}{%
{\protect \APACyear {1995}}%
}]{%
harris1995influence}
\APACinsertmetastar {%
harris1995influence}%
\begin{APACrefauthors}%
Harris, R\BPBI A.%
, Simpson, R\BPBI W.%
\BCBL {}\ \BBA {} Reasenberg, P\BPBI A.%
\end{APACrefauthors}%
\unskip\
\newblock
\APACrefYearMonthDay{1995}{}{}.
\newblock
{\BBOQ}\APACrefatitle {{Influence of static stress changes on earthquake locations in southern California}} {{Influence of static stress changes on earthquake locations in southern California}}.{\BBCQ}
\newblock
\APACjournalVolNumPages{Nature}{375}{6528}{221--224}.
\PrintBackRefs{\CurrentBib}

\bibitem [\protect \citeauthoryear {%
Hauksson%
}{%
Hauksson%
}{%
{\protect \APACyear {1987}}%
}]{%
hauksson1987seismotectonics}
\APACinsertmetastar {%
hauksson1987seismotectonics}%
\begin{APACrefauthors}%
Hauksson, E.%
\end{APACrefauthors}%
\unskip\
\newblock
\APACrefYearMonthDay{1987}{}{}.
\newblock
{\BBOQ}\APACrefatitle {{Seismotectonics of the Newport-Inglewood fault zone in the Los Angeles basin, southern California}} {{Seismotectonics of the Newport-Inglewood fault zone in the Los Angeles basin, southern California}}.{\BBCQ}
\newblock
\APACjournalVolNumPages{Bulletin of the Seismological Society of America}{77}{2}{539--561}.
\PrintBackRefs{\CurrentBib}

\bibitem [\protect \citeauthoryear {%
Hauksson%
}{%
Hauksson%
}{%
{\protect \APACyear {1990}}%
}]{%
hauksson1990earthquakes}
\APACinsertmetastar {%
hauksson1990earthquakes}%
\begin{APACrefauthors}%
Hauksson, E.%
\end{APACrefauthors}%
\unskip\
\newblock
\APACrefYearMonthDay{1990}{}{}.
\newblock
{\BBOQ}\APACrefatitle {{Earthquakes, faulting, and stress in the Los Angeles basin}} {{Earthquakes, faulting, and stress in the Los Angeles basin}}.{\BBCQ}
\newblock
\APACjournalVolNumPages{Journal of Geophysical Research: Solid Earth}{95}{B10}{15365--15394}.
\PrintBackRefs{\CurrentBib}

\bibitem [\protect \citeauthoryear {%
Hauksson%
\ \BBA {} Gross%
}{%
Hauksson%
\ \BBA {} Gross%
}{%
{\protect \APACyear {1991}}%
}]{%
hauksson1991source}
\APACinsertmetastar {%
hauksson1991source}%
\begin{APACrefauthors}%
Hauksson, E.%
\BCBT {}\ \BBA {} Gross, S.%
\end{APACrefauthors}%
\unskip\
\newblock
\APACrefYearMonthDay{1991}{}{}.
\newblock
{\BBOQ}\APACrefatitle {{Source parameters of the 1933 Long Beach earthquake}} {{Source parameters of the 1933 Long Beach earthquake}}.{\BBCQ}
\newblock
\APACjournalVolNumPages{Bulletin of the Seismological Society of America}{81}{1}{81--98}.
\PrintBackRefs{\CurrentBib}

\bibitem [\protect \citeauthoryear {%
Hauksson%
, Yang%
\BCBL {}\ \BBA {} Shearer%
}{%
Hauksson%
\ \protect \BOthers {.}}{%
{\protect \APACyear {2012}}%
}]{%
hauksson2012waveform}
\APACinsertmetastar {%
hauksson2012waveform}%
\begin{APACrefauthors}%
Hauksson, E.%
, Yang, W.%
\BCBL {}\ \BBA {} Shearer, P\BPBI M.%
\end{APACrefauthors}%
\unskip\
\newblock
\APACrefYearMonthDay{2012}{}{}.
\newblock
{\BBOQ}\APACrefatitle {{Waveform relocated earthquake catalog for southern California (1981 to June 2011)}} {{Waveform relocated earthquake catalog for southern California (1981 to June 2011)}}.{\BBCQ}
\newblock
\APACjournalVolNumPages{Bulletin of the Seismological Society of America}{102}{5}{2239--2244}.
\PrintBackRefs{\CurrentBib}

\bibitem [\protect \citeauthoryear {%
Herrmann%
}{%
Herrmann%
}{%
{\protect \APACyear {2013}}%
}]{%
herrmann2013computer}
\APACinsertmetastar {%
herrmann2013computer}%
\begin{APACrefauthors}%
Herrmann, R\BPBI B.%
\end{APACrefauthors}%
\unskip\
\newblock
\APACrefYearMonthDay{2013}{}{}.
\newblock
{\BBOQ}\APACrefatitle {{Computer programs in seismology: An evolving tool for instruction and research}} {{Computer programs in seismology: An evolving tool for instruction and research}}.{\BBCQ}
\newblock
\APACjournalVolNumPages{Seismological Research Letters}{84}{6}{1081--1088}.
\PrintBackRefs{\CurrentBib}

\bibitem [\protect \citeauthoryear {%
Hough%
\ \BBA {} Graves%
}{%
Hough%
\ \BBA {} Graves%
}{%
{\protect \APACyear {2020}}%
}]{%
hough20201933}
\APACinsertmetastar {%
hough20201933}%
\begin{APACrefauthors}%
Hough, S\BPBI E.%
\BCBT {}\ \BBA {} Graves, R.%
\end{APACrefauthors}%
\unskip\
\newblock
\APACrefYearMonthDay{2020}{}{}.
\newblock
{\BBOQ}\APACrefatitle {{The 1933 Long Beach earthquake (California, USA): Ground motions and rupture scenario}} {{The 1933 Long Beach earthquake (California, USA): Ground motions and rupture scenario}}.{\BBCQ}
\newblock
\APACjournalVolNumPages{Scientific Reports}{10}{1}{10017}.
\PrintBackRefs{\CurrentBib}

\bibitem [\protect \citeauthoryear {%
Jia%
\ \BBA {} Clayton%
}{%
Jia%
\ \BBA {} Clayton%
}{%
{\protect \APACyear {2021}}%
}]{%
jia2021determination}
\APACinsertmetastar {%
jia2021determination}%
\begin{APACrefauthors}%
Jia, Z.%
\BCBT {}\ \BBA {} Clayton, R\BPBI W.%
\end{APACrefauthors}%
\unskip\
\newblock
\APACrefYearMonthDay{2021}{}{}.
\newblock
{\BBOQ}\APACrefatitle {{Determination of near surface shear-wave velocities in the central Los Angeles basin with dense arrays}} {{Determination of near surface shear-wave velocities in the central Los Angeles basin with dense arrays}}.{\BBCQ}
\newblock
\APACjournalVolNumPages{Journal of Geophysical Research: Solid Earth}{126}{5}{e2020JB021369}.
\PrintBackRefs{\CurrentBib}

\bibitem [\protect \citeauthoryear {%
Jin%
\ \BBA {} Kim%
}{%
Jin%
\ \BBA {} Kim%
}{%
{\protect \APACyear {2021}}%
}]{%
jin2021importance}
\APACinsertmetastar {%
jin2021importance}%
\begin{APACrefauthors}%
Jin, K.%
\BCBT {}\ \BBA {} Kim, Y\BHBI S.%
\end{APACrefauthors}%
\unskip\
\newblock
\APACrefYearMonthDay{2021}{}{}.
\newblock
{\BBOQ}\APACrefatitle {The importance of surface ruptures and fault damage zones in earthquake hazard assessment: a review and new suggestions} {The importance of surface ruptures and fault damage zones in earthquake hazard assessment: a review and new suggestions}.{\BBCQ}
\newblock
\BIn{} Y.~Okubo\ (\BED), \APACrefbtitle {Characterization of Modern and Historical Seismic–Tsunamic Events, and Their Global–Societal Impacts.} {Characterization of modern and historical seismic–tsunamic events, and their global–societal impacts.}
\newblock
\APACaddressPublisher{London}{Geological Society of London}.
\PrintBackRefs{\CurrentBib}

\bibitem [\protect \citeauthoryear {%
Y.~Li%
}{%
Y.~Li%
}{%
{\protect \APACyear {2023}}%
}]{%
yi2023thesis}
\APACinsertmetastar {%
yi2023thesis}%
\begin{APACrefauthors}%
Li, Y.%
\end{APACrefauthors}%
\unskip\
\newblock
\APACrefYear{2023}.
\unskip\
\newblock
\APACrefbtitle {Dynamics of Subduction Initiation and Constraining Sedimentary Basin Structure with Seismic Ambient Noise} {Dynamics of subduction initiation and constraining sedimentary basin structure with seismic ambient noise}\ \APACtypeAddressSchool {\BUPhD}{}{}.
\unskip\
\newblock
\APACaddressSchool {}{California Institute of Technology}.
\PrintBackRefs{\CurrentBib}

\bibitem [\protect \citeauthoryear {%
Y\BHBI G.~Li%
\ \BBA {} Leary%
}{%
Y\BHBI G.~Li%
\ \BBA {} Leary%
}{%
{\protect \APACyear {1990}}%
}]{%
li1990fault}
\APACinsertmetastar {%
li1990fault}%
\begin{APACrefauthors}%
Li, Y\BHBI G.%
\BCBT {}\ \BBA {} Leary, P.%
\end{APACrefauthors}%
\unskip\
\newblock
\APACrefYearMonthDay{1990}{}{}.
\newblock
{\BBOQ}\APACrefatitle {Fault zone trapped seismic waves} {Fault zone trapped seismic waves}.{\BBCQ}
\newblock
\APACjournalVolNumPages{Bulletin of the Seismological Society of America}{80}{5}{1245--1271}.
\PrintBackRefs{\CurrentBib}

\bibitem [\protect \citeauthoryear {%
Lin%
, Li%
, Clayton%
\BCBL {}\ \BBA {} Hollis%
}{%
Lin%
\ \protect \BOthers {.}}{%
{\protect \APACyear {2013}}%
}]{%
lin2013high}
\APACinsertmetastar {%
lin2013high}%
\begin{APACrefauthors}%
Lin, F\BHBI C.%
, Li, D.%
, Clayton, R\BPBI W.%
\BCBL {}\ \BBA {} Hollis, D.%
\end{APACrefauthors}%
\unskip\
\newblock
\APACrefYearMonthDay{2013}{}{}.
\newblock
{\BBOQ}\APACrefatitle {{High-resolution 3D shallow crustal structure in Long Beach, California: Application of ambient noise tomography on a dense seismic array}} {{High-resolution 3D shallow crustal structure in Long Beach, California: Application of ambient noise tomography on a dense seismic array}}.{\BBCQ}
\newblock
\APACjournalVolNumPages{Geophysics}{78}{4}{Q45--Q56}.
\PrintBackRefs{\CurrentBib}

\bibitem [\protect \citeauthoryear {%
Liu%
, Beroza%
, Yang%
\BCBL {}\ \BBA {} Ellsworth%
}{%
Liu%
\ \protect \BOthers {.}}{%
{\protect \APACyear {2021}}%
}]{%
liu2021ambient}
\APACinsertmetastar {%
liu2021ambient}%
\begin{APACrefauthors}%
Liu, X.%
, Beroza, G\BPBI C.%
, Yang, L.%
\BCBL {}\ \BBA {} Ellsworth, W\BPBI L.%
\end{APACrefauthors}%
\unskip\
\newblock
\APACrefYearMonthDay{2021}{}{}.
\newblock
{\BBOQ}\APACrefatitle {{Ambient noise Love wave attenuation tomography for the LASSIE array across the Los Angeles basin}} {{Ambient noise Love wave attenuation tomography for the LASSIE array across the Los Angeles basin}}.{\BBCQ}
\newblock
\APACjournalVolNumPages{Science Advances}{7}{22}{eabe1030}.
\PrintBackRefs{\CurrentBib}

\bibitem [\protect \citeauthoryear {%
Ma%
, Clayton%
, Tsai%
\BCBL {}\ \BBA {} Zhan%
}{%
Ma%
\ \protect \BOthers {.}}{%
{\protect \APACyear {2013}}%
}]{%
ma2013locating}
\APACinsertmetastar {%
ma2013locating}%
\begin{APACrefauthors}%
Ma, Y.%
, Clayton, R\BPBI W.%
, Tsai, V\BPBI C.%
\BCBL {}\ \BBA {} Zhan, Z.%
\end{APACrefauthors}%
\unskip\
\newblock
\APACrefYearMonthDay{2013}{}{}.
\newblock
{\BBOQ}\APACrefatitle {{Locating a scatterer in the active volcanic area of Southern Peru from ambient noise cross-correlation}} {{Locating a scatterer in the active volcanic area of Southern Peru from ambient noise cross-correlation}}.{\BBCQ}
\newblock
\APACjournalVolNumPages{Geophysical Journal International}{192}{3}{1332--1341}.
\PrintBackRefs{\CurrentBib}

\bibitem [\protect \citeauthoryear {%
Nakata%
, Chang%
, Lawrence%
\BCBL {}\ \BBA {} Bou{\'e}%
}{%
Nakata%
\ \protect \BOthers {.}}{%
{\protect \APACyear {2015}}%
}]{%
nakata2015body}
\APACinsertmetastar {%
nakata2015body}%
\begin{APACrefauthors}%
Nakata, N.%
, Chang, J\BPBI P.%
, Lawrence, J\BPBI F.%
\BCBL {}\ \BBA {} Bou{\'e}, P.%
\end{APACrefauthors}%
\unskip\
\newblock
\APACrefYearMonthDay{2015}{}{}.
\newblock
{\BBOQ}\APACrefatitle {{Body wave extraction and tomography at Long Beach, California, with ambient-noise interferometry}} {{Body wave extraction and tomography at Long Beach, California, with ambient-noise interferometry}}.{\BBCQ}
\newblock
\APACjournalVolNumPages{Journal of Geophysical Research: Solid Earth}{120}{2}{1159--1173}.
\PrintBackRefs{\CurrentBib}

\bibitem [\protect \citeauthoryear {%
Olsen%
\ \protect \BOthers {.}}{%
Olsen%
\ \protect \BOthers {.}}{%
{\protect \APACyear {2009}}%
}]{%
olsen2009shakeout}
\APACinsertmetastar {%
olsen2009shakeout}%
\begin{APACrefauthors}%
Olsen, K.%
, Day, S.%
, Dalguer, L.%
, Mayhew, J.%
, Cui, Y.%
, Zhu, J.%
\BDBL {}others%
\end{APACrefauthors}%
\unskip\
\newblock
\APACrefYearMonthDay{2009}{}{}.
\newblock
{\BBOQ}\APACrefatitle {{S}hake{O}ut-{D}: {G}round motion estimates using an ensemble of large earthquakes on the southern {S}an {A}ndreas fault with spontaneous rupture propagation} {{S}hake{O}ut-{D}: {G}round motion estimates using an ensemble of large earthquakes on the southern {S}an {A}ndreas fault with spontaneous rupture propagation}.{\BBCQ}
\newblock
\APACjournalVolNumPages{Geophysical Research Letters}{36}{4}{}.
\PrintBackRefs{\CurrentBib}

\bibitem [\protect \citeauthoryear {%
Olsen%
\ \protect \BOthers {.}}{%
Olsen%
\ \protect \BOthers {.}}{%
{\protect \APACyear {2006}}%
}]{%
olsen2006strong}
\APACinsertmetastar {%
olsen2006strong}%
\begin{APACrefauthors}%
Olsen, K.%
, Day, S.%
, Minster, J.%
, Cui, Y.%
, Chourasia, A.%
, Faerman, M.%
\BDBL {}Jordan, T.%
\end{APACrefauthors}%
\unskip\
\newblock
\APACrefYearMonthDay{2006}{}{}.
\newblock
{\BBOQ}\APACrefatitle {Strong shaking in {L}os {A}ngeles expected from southern {S}an {A}ndreas earthquake} {Strong shaking in {L}os {A}ngeles expected from southern {S}an {A}ndreas earthquake}.{\BBCQ}
\newblock
\APACjournalVolNumPages{Geophysical Research Letters}{33}{7}{}.
\PrintBackRefs{\CurrentBib}

\bibitem [\protect \citeauthoryear {%
Reitherman%
}{%
Reitherman%
}{%
{\protect \APACyear {1992}}%
}]{%
reitherman1992effectiveness}
\APACinsertmetastar {%
reitherman1992effectiveness}%
\begin{APACrefauthors}%
Reitherman, R.%
\end{APACrefauthors}%
\unskip\
\newblock
\APACrefYearMonthDay{1992}{}{}.
\newblock
{\BBOQ}\APACrefatitle {{The Effectiveness of Fault Zone Regulations in California}} {{The Effectiveness of Fault Zone Regulations in California}}.{\BBCQ}
\newblock
\APACjournalVolNumPages{Earthquake Spectra}{8}{1}{57--77}.
\PrintBackRefs{\CurrentBib}

\bibitem [\protect \citeauthoryear {%
Retailleau%
, Bou{\'e}%
, Stehly%
\BCBL {}\ \BBA {} Campillo%
}{%
Retailleau%
\ \protect \BOthers {.}}{%
{\protect \APACyear {2017}}%
}]{%
retailleau2017locating}
\APACinsertmetastar {%
retailleau2017locating}%
\begin{APACrefauthors}%
Retailleau, L.%
, Bou{\'e}, P.%
, Stehly, L.%
\BCBL {}\ \BBA {} Campillo, M.%
\end{APACrefauthors}%
\unskip\
\newblock
\APACrefYearMonthDay{2017}{}{}.
\newblock
{\BBOQ}\APACrefatitle {Locating microseism sources using spurious arrivals in intercontinental noise correlations} {Locating microseism sources using spurious arrivals in intercontinental noise correlations}.{\BBCQ}
\newblock
\APACjournalVolNumPages{Journal of Geophysical Research: Solid Earth}{122}{10}{8107--8120}.
\PrintBackRefs{\CurrentBib}

\bibitem [\protect \citeauthoryear {%
Robertsson%
}{%
Robertsson%
}{%
{\protect \APACyear {1996}}%
}]{%
robertsson1996numerical}
\APACinsertmetastar {%
robertsson1996numerical}%
\begin{APACrefauthors}%
Robertsson, J\BPBI O.%
\end{APACrefauthors}%
\unskip\
\newblock
\APACrefYearMonthDay{1996}{}{}.
\newblock
{\BBOQ}\APACrefatitle {A numerical free-surface condition for elastic/viscoelastic finite-difference modeling in the presence of topography} {A numerical free-surface condition for elastic/viscoelastic finite-difference modeling in the presence of topography}.{\BBCQ}
\newblock
\APACjournalVolNumPages{Geophysics}{61}{6}{1921--1934}.
\PrintBackRefs{\CurrentBib}

\bibitem [\protect \citeauthoryear {%
Ross%
, Trugman%
, Hauksson%
\BCBL {}\ \BBA {} Shearer%
}{%
Ross%
\ \protect \BOthers {.}}{%
{\protect \APACyear {2019}}%
}]{%
ross2019searching}
\APACinsertmetastar {%
ross2019searching}%
\begin{APACrefauthors}%
Ross, Z\BPBI E.%
, Trugman, D\BPBI T.%
, Hauksson, E.%
\BCBL {}\ \BBA {} Shearer, P\BPBI M.%
\end{APACrefauthors}%
\unskip\
\newblock
\APACrefYearMonthDay{2019}{}{}.
\newblock
{\BBOQ}\APACrefatitle {Searching for hidden earthquakes in {S}outhern {C}alifornia} {Searching for hidden earthquakes in {S}outhern {C}alifornia}.{\BBCQ}
\newblock
\APACjournalVolNumPages{Science}{364}{6442}{767--771}.
\PrintBackRefs{\CurrentBib}

\bibitem [\protect \citeauthoryear {%
Shapiro%
\ \BBA {} Campillo%
}{%
Shapiro%
\ \BBA {} Campillo%
}{%
{\protect \APACyear {2004}}%
}]{%
shapiro2004emergence}
\APACinsertmetastar {%
shapiro2004emergence}%
\begin{APACrefauthors}%
Shapiro, N\BPBI M.%
\BCBT {}\ \BBA {} Campillo, M.%
\end{APACrefauthors}%
\unskip\
\newblock
\APACrefYearMonthDay{2004}{}{}.
\newblock
{\BBOQ}\APACrefatitle {{Emergence of broadband Rayleigh waves from correlations of the ambient seismic noise}} {{Emergence of broadband Rayleigh waves from correlations of the ambient seismic noise}}.{\BBCQ}
\newblock
\APACjournalVolNumPages{Geophysical Research Letters}{31}{7}{}.
\PrintBackRefs{\CurrentBib}

\bibitem [\protect \citeauthoryear {%
Shaw%
\ \BBA {} Suppe%
}{%
Shaw%
\ \BBA {} Suppe%
}{%
{\protect \APACyear {1996}}%
}]{%
shaw1996earthquake}
\APACinsertmetastar {%
shaw1996earthquake}%
\begin{APACrefauthors}%
Shaw, J\BPBI H.%
\BCBT {}\ \BBA {} Suppe, J.%
\end{APACrefauthors}%
\unskip\
\newblock
\APACrefYearMonthDay{1996}{}{}.
\newblock
{\BBOQ}\APACrefatitle {{Earthquake hazards of active blind-thrust faults under the central Los Angeles basin, California}} {{Earthquake hazards of active blind-thrust faults under the central Los Angeles basin, California}}.{\BBCQ}
\newblock
\APACjournalVolNumPages{Journal of Geophysical Research: Solid Earth}{101}{B4}{8623--8642}.
\PrintBackRefs{\CurrentBib}

\bibitem [\protect \citeauthoryear {%
Socco%
\ \BBA {} Strobbia%
}{%
Socco%
\ \BBA {} Strobbia%
}{%
{\protect \APACyear {2004}}%
}]{%
socco2004surface}
\APACinsertmetastar {%
socco2004surface}%
\begin{APACrefauthors}%
Socco, L.%
\BCBT {}\ \BBA {} Strobbia, C.%
\end{APACrefauthors}%
\unskip\
\newblock
\APACrefYearMonthDay{2004}{}{}.
\newblock
{\BBOQ}\APACrefatitle {Surface-wave method for near-surface characterization: A tutorial} {Surface-wave method for near-surface characterization: A tutorial}.{\BBCQ}
\newblock
\APACjournalVolNumPages{Near surface geophysics}{2}{4}{165--185}.
\PrintBackRefs{\CurrentBib}

\bibitem [\protect \citeauthoryear {%
Storchak%
\ \protect \BOthers {.}}{%
Storchak%
\ \protect \BOthers {.}}{%
{\protect \APACyear {2013}}%
}]{%
storchak2013public}
\APACinsertmetastar {%
storchak2013public}%
\begin{APACrefauthors}%
Storchak, D\BPBI A.%
, Di~Giacomo, D.%
, Bond{\'a}r, I.%
, Engdahl, E\BPBI R.%
, Harris, J.%
, Lee, W\BPBI H.%
\BDBL {}Bormann, P.%
\end{APACrefauthors}%
\unskip\
\newblock
\APACrefYearMonthDay{2013}{}{}.
\newblock
{\BBOQ}\APACrefatitle {Public release of the {ISC--GEM} global instrumental earthquake catalogue (1900--2009)} {Public release of the {ISC--GEM} global instrumental earthquake catalogue (1900--2009)}.{\BBCQ}
\newblock
\APACjournalVolNumPages{Seismological Research Letters}{84}{5}{810--815}.
\PrintBackRefs{\CurrentBib}

\bibitem [\protect \citeauthoryear {%
Taber%
}{%
Taber%
}{%
{\protect \APACyear {1920}}%
}]{%
taber1920inglewood}
\APACinsertmetastar {%
taber1920inglewood}%
\begin{APACrefauthors}%
Taber, S.%
\end{APACrefauthors}%
\unskip\
\newblock
\APACrefYearMonthDay{1920}{}{}.
\newblock
{\BBOQ}\APACrefatitle {{The Inglewood earthquake in southern California, June 21, 1920}} {{The Inglewood earthquake in southern California, June 21, 1920}}.{\BBCQ}
\newblock
\APACjournalVolNumPages{Bulletin of the Seismological Society of America}{10}{3}{129--145}.
\PrintBackRefs{\CurrentBib}

\bibitem [\protect \citeauthoryear {%
Wald%
\ \BBA {} Graves%
}{%
Wald%
\ \BBA {} Graves%
}{%
{\protect \APACyear {1998}}%
}]{%
wald1998seismic}
\APACinsertmetastar {%
wald1998seismic}%
\begin{APACrefauthors}%
Wald, D\BPBI J.%
\BCBT {}\ \BBA {} Graves, R\BPBI W.%
\end{APACrefauthors}%
\unskip\
\newblock
\APACrefYearMonthDay{1998}{}{}.
\newblock
{\BBOQ}\APACrefatitle {The seismic response of the {L}os {A}ngeles basin, {C}alifornia} {The seismic response of the {L}os {A}ngeles basin, {C}alifornia}.{\BBCQ}
\newblock
\APACjournalVolNumPages{{B}ulletin of the {S}eismological {S}ociety of {A}merica}{88}{2}{337--356}.
\PrintBackRefs{\CurrentBib}

\bibitem [\protect \citeauthoryear {%
Wilson%
}{%
Wilson%
}{%
{\protect \APACyear {2015}}%
}]{%
wilson2015characterization}
\APACinsertmetastar {%
wilson2015characterization}%
\begin{APACrefauthors}%
Wilson, G.%
\end{APACrefauthors}%
\unskip\
\newblock
\APACrefYear{2015}.
\newblock
\APACrefbtitle {{Characterization of the Alamitos Heights Fault beneath California State University, Long Beach: A splay of the Newport-Inglewood fault zone}} {{Characterization of the Alamitos Heights Fault beneath California State University, Long Beach: A splay of the Newport-Inglewood fault zone}}.
\newblock
\APACaddressPublisher{}{California State University, Long Beach}.
\PrintBackRefs{\CurrentBib}

\bibitem [\protect \citeauthoryear {%
Wood%
}{%
Wood%
}{%
{\protect \APACyear {1933}}%
}]{%
wood1933long}
\APACinsertmetastar {%
wood1933long}%
\begin{APACrefauthors}%
Wood, H\BPBI O.%
\end{APACrefauthors}%
\unskip\
\newblock
\APACrefYearMonthDay{1933}{}{}.
\newblock
{\BBOQ}\APACrefatitle {The long beach earthquake} {The long beach earthquake}.{\BBCQ}
\newblock
\APACjournalVolNumPages{Science}{78}{2016}{147--148}.
\PrintBackRefs{\CurrentBib}

\bibitem [\protect \citeauthoryear {%
Yang%
\ \BBA {} Clayton%
}{%
Yang%
\ \BBA {} Clayton%
}{%
{\protect \APACyear {2023}}%
}]{%
yang2023shallow}
\APACinsertmetastar {%
yang2023shallow}%
\begin{APACrefauthors}%
Yang, Y.%
\BCBT {}\ \BBA {} Clayton, R\BPBI W.%
\end{APACrefauthors}%
\unskip\
\newblock
\APACrefYearMonthDay{2023}{}{}.
\newblock
{\BBOQ}\APACrefatitle {{Shallow Seismicity in the Long Beach--Seal Beach, California Area}} {{Shallow Seismicity in the Long Beach--Seal Beach, California Area}}.{\BBCQ}
\newblock
\APACjournalVolNumPages{Seismological Society of America}{}{}{}.
\PrintBackRefs{\CurrentBib}

\bibitem [\protect \citeauthoryear {%
Yao%
, Gouedard%
, Collins%
, McGuire%
\BCBL {}\ \BBA {} van~der Hilst%
}{%
Yao%
\ \protect \BOthers {.}}{%
{\protect \APACyear {2011}}%
}]{%
yao2011structure}
\APACinsertmetastar {%
yao2011structure}%
\begin{APACrefauthors}%
Yao, H.%
, Gouedard, P.%
, Collins, J\BPBI A.%
, McGuire, J\BPBI J.%
\BCBL {}\ \BBA {} van~der Hilst, R\BPBI D.%
\end{APACrefauthors}%
\unskip\
\newblock
\APACrefYearMonthDay{2011}{}{}.
\newblock
{\BBOQ}\APACrefatitle {{Structure of young East Pacific Rise lithosphere from ambient noise correlation analysis of fundamental-and higher-mode Scholte-Rayleigh waves}} {{Structure of young East Pacific Rise lithosphere from ambient noise correlation analysis of fundamental-and higher-mode Scholte-Rayleigh waves}}.{\BBCQ}
\newblock
\APACjournalVolNumPages{Comptes Rendus Geoscience}{343}{8-9}{571--583}.
\PrintBackRefs{\CurrentBib}

\bibitem [\protect \citeauthoryear {%
Yeats%
}{%
Yeats%
}{%
{\protect \APACyear {1973}}%
}]{%
yeats1973newport}
\APACinsertmetastar {%
yeats1973newport}%
\begin{APACrefauthors}%
Yeats, R\BPBI S.%
\end{APACrefauthors}%
\unskip\
\newblock
\APACrefYearMonthDay{1973}{}{}.
\newblock
{\BBOQ}\APACrefatitle {{Newport-Inglewood fault zone, Los Angeles basin, California}} {{Newport-Inglewood fault zone, Los Angeles basin, California}}.{\BBCQ}
\newblock
\APACjournalVolNumPages{AAPG Bulletin}{57}{1}{117--135}.
\PrintBackRefs{\CurrentBib}

\end{thebibliography}

\section*{Acknowledgements}

We gratefully acknowledge LA seismic (LAseismic.com) for the use of the Seal Beach data. We thank Eric Campbell for facilitating the use of the seismic data and Dan Gish and Steve Boljen for providing us with the migrated seismic sections of the Seal Beach survey. 

\section*{Author contributions statement}

Conceptualization: EB, RWC
Methodology: EB, JCC, RWC
Visualization: EB
Supervision: RWC
Writing—original draft: EB
Writing—review \& editing: EB, JCC, RWC

\subsection*{Contact information}
\begin{itemize}
    \item {\bf{Ettore Biondi}}: California Institute of Technology, Seismological Laboratory; RM 262; 1200 E. California Blvd., MS 252-21 Pasadena, California 91125-2100
    \item {\bf{Jorge C. Castellanos}}: Risk Management Solutions; 7575 Gateway Blvd., Suite 300, Newark, CA 94560, USA
    \item {\bf{Robert W. Clayton}}: California Institute of Technology, Seismological Laboratory; RM 352; 1200 E. California Blvd., MS 252-21 Pasadena, California 91125-2100
\end{itemize}

\section*{Data and Resources}
Permission from LA Seismic is required to access the raw seismic data. The correlograms computed from the raw data and employed in this study will be uploaded on Zenodo upon acceptance of the manuscript.

\section*{Additional information}

\subsection*{Competing interests}
The authors declare no competing interests.

\subsection*{Supplementary materials}
Supplementary Figures 1 and 2 display other comparisons of our results with the active survey profiles. Supplementary Figure 3 shows the full seismograms of the synthetic depth-sensitivity test; while Supplementary Figure 4 depicts our results on the Seal Beach array data with different processing parameters. The two Supplementary movies show how the different arrivals appear within noise cross-correlation linear arrays.

\begin{figure}[ht]
\centering
\includegraphics[width=0.7\linewidth]{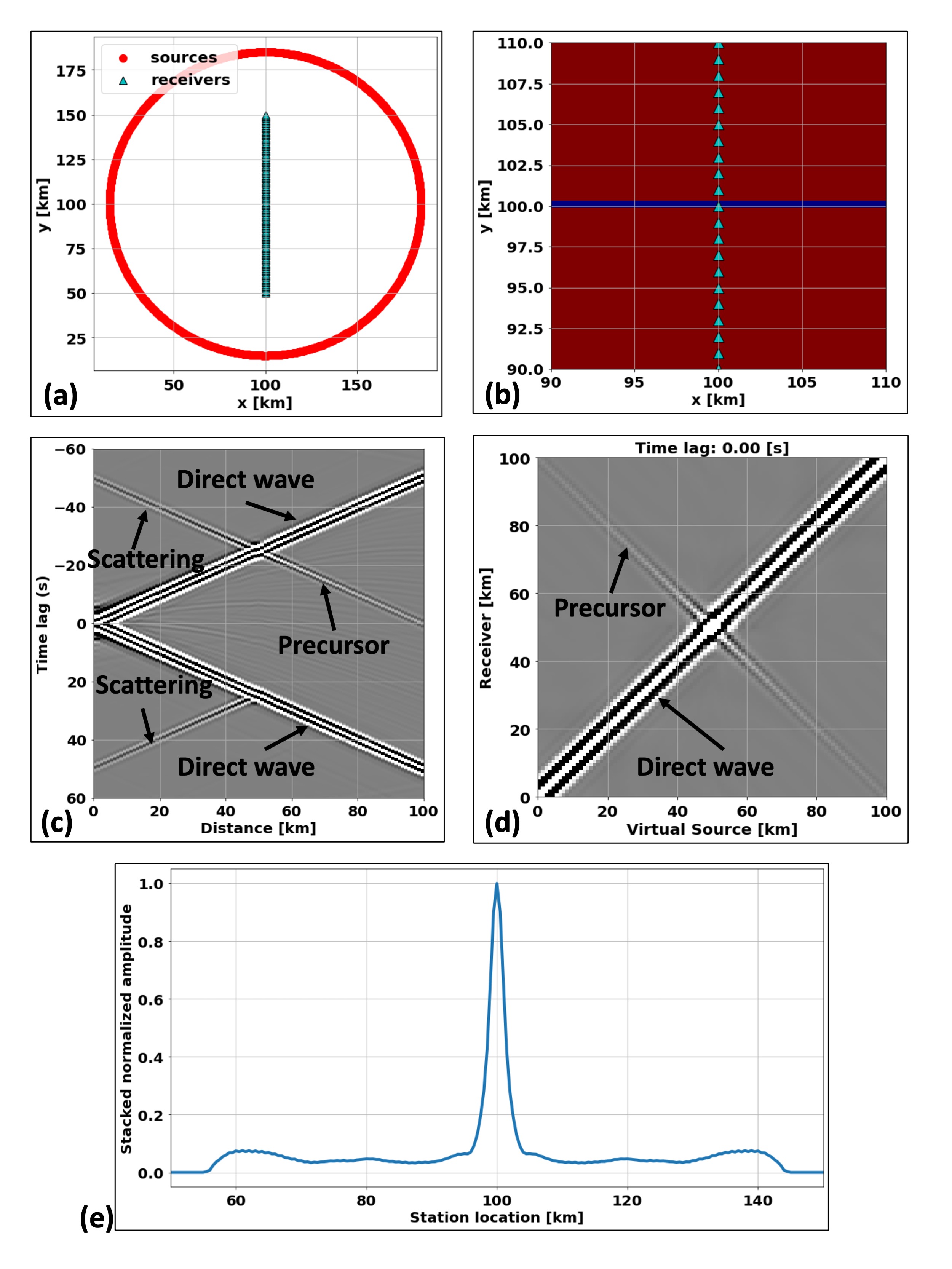}
\caption{Synthetic precursor imaging test. (a) Receiver and sources considered in the synthetic cross-correlation test. (b) Velocity model employed in the synthetic precursor test. (c) Representative cross-correlation for a virtual source highlighting the direct wave and precursory energy. (d) Zero-time lag slice from the node line used in the synthetic test. (e) Stacked precursory energy for the synthetic test. The peak in energy corresponds to the location of the increase in the velocity model causing the recorded scattering.}
\label{fig:figure6}
\end{figure}

\begin{figure}[ht]
\centering
\includegraphics[width=0.7\linewidth]{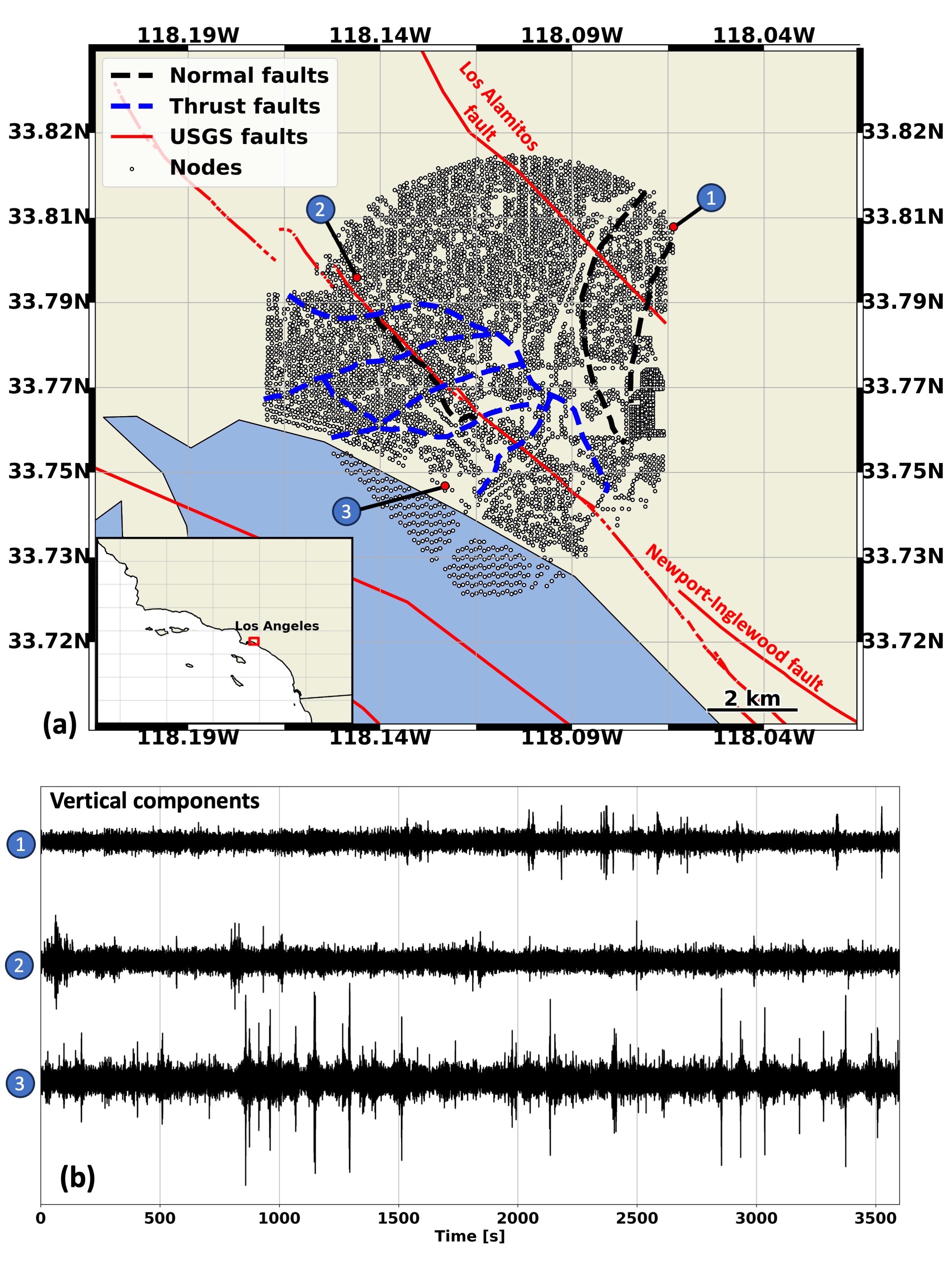}
\caption{Study area and location of the Seal Beach seismic dense array. (a) Map showing the location of the Seal Beach nodal array (white dots) with highlighted the mapped faults~\cite{frankel2000usgs} (red, blue, and black lines). The map inset shows the location of the area (red rectangle) within Southern California. (b) Example vertical-component seismograms for the three nodes indicated in (a).}
\label{fig:figure1}
\end{figure}

\begin{figure}[ht]
\centering
\includegraphics[width=0.9\linewidth]{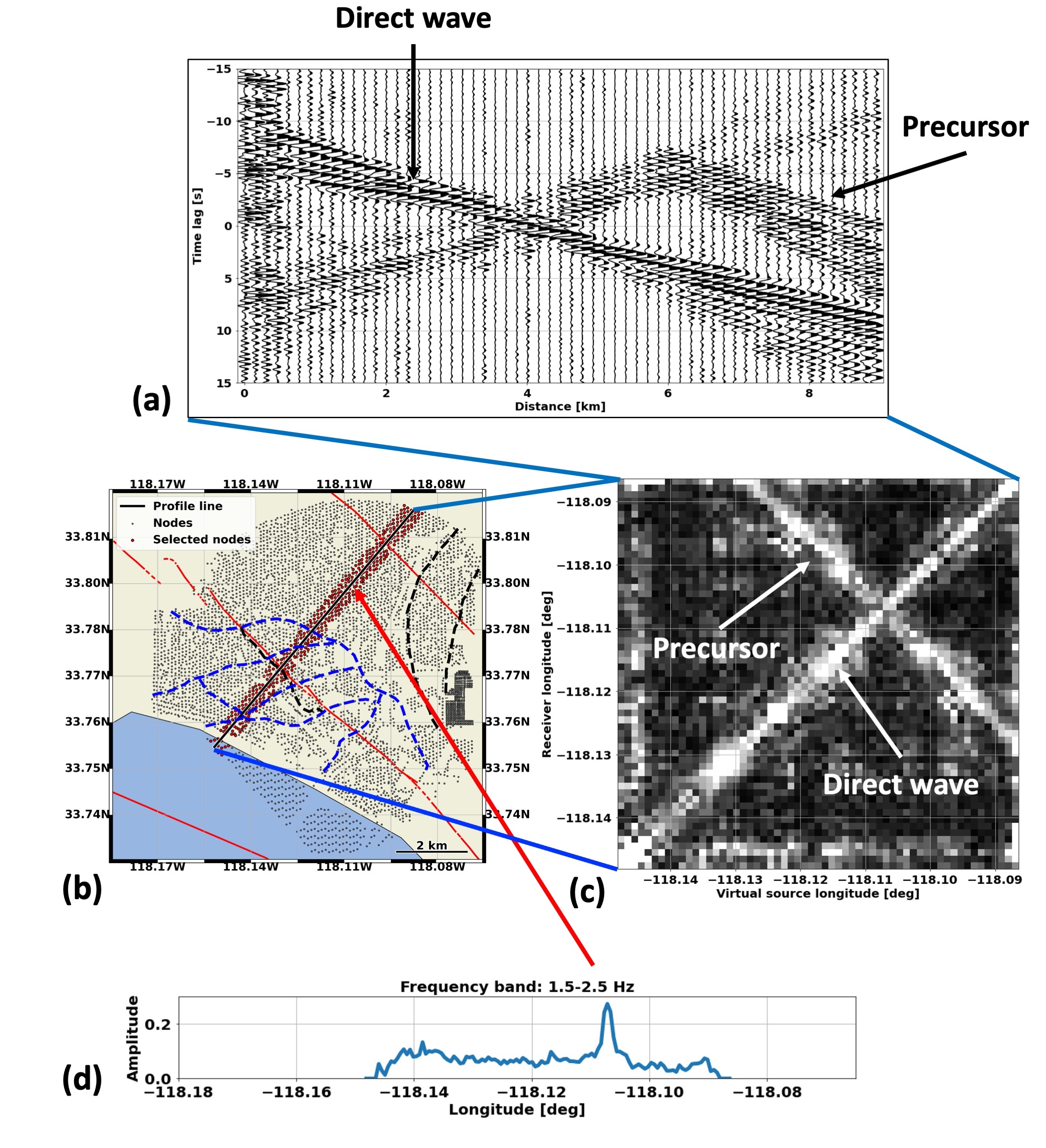}
\caption{Precursory energy observation on the Seal Beach dense array. (a) Representative binned noise cross-correlation for a given source along the highlighted nodes in panel B. (b) Location of the nodes considered for the binning procedure (red dots). The black solid line depicts the source-receiver line used for the binning process. (c) Zero-time-lag slice of the envelope of the CCs extracted from the source-receiver-time cross-correlation cube in which the direct and precursory arrivals are indicated.  (d) Stacked precursory energy obtained from this line of nodes. The red arrow indicates the location of the peak obtained from the stacked energy. }
\label{fig:figure2}
\end{figure}

\begin{figure}[ht]
\centering
\includegraphics[width=0.9\linewidth]{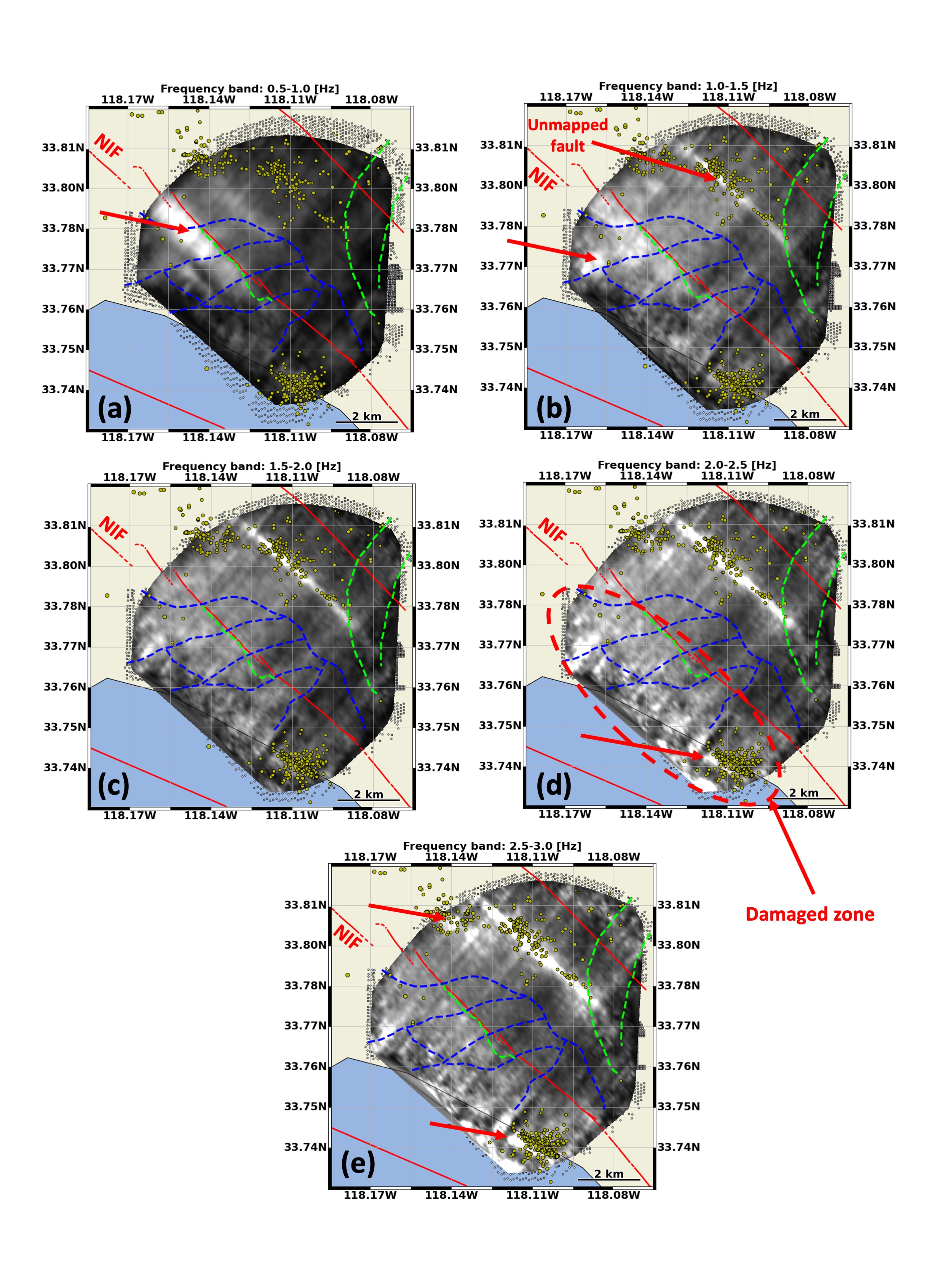}
\caption{Precursor stacked images at various frequencies. (a-b) Images obtained by processing various node lines with cross-correlation bandpassed at the frequencies shown about each panel. The red lines indicate the USGS-mapped faults, while the blue and green lines are mapped from the active survey images. The yellow dots depict the shallow earthquakes cataloged by a previous study~\cite{yang2023shallow}. The red arrows indicate relevant scattering features obtained from the described imaging procedure.}
\label{fig:figure3}
\end{figure}

\begin{figure}[ht]
\centering
\includegraphics[width=0.9\linewidth]{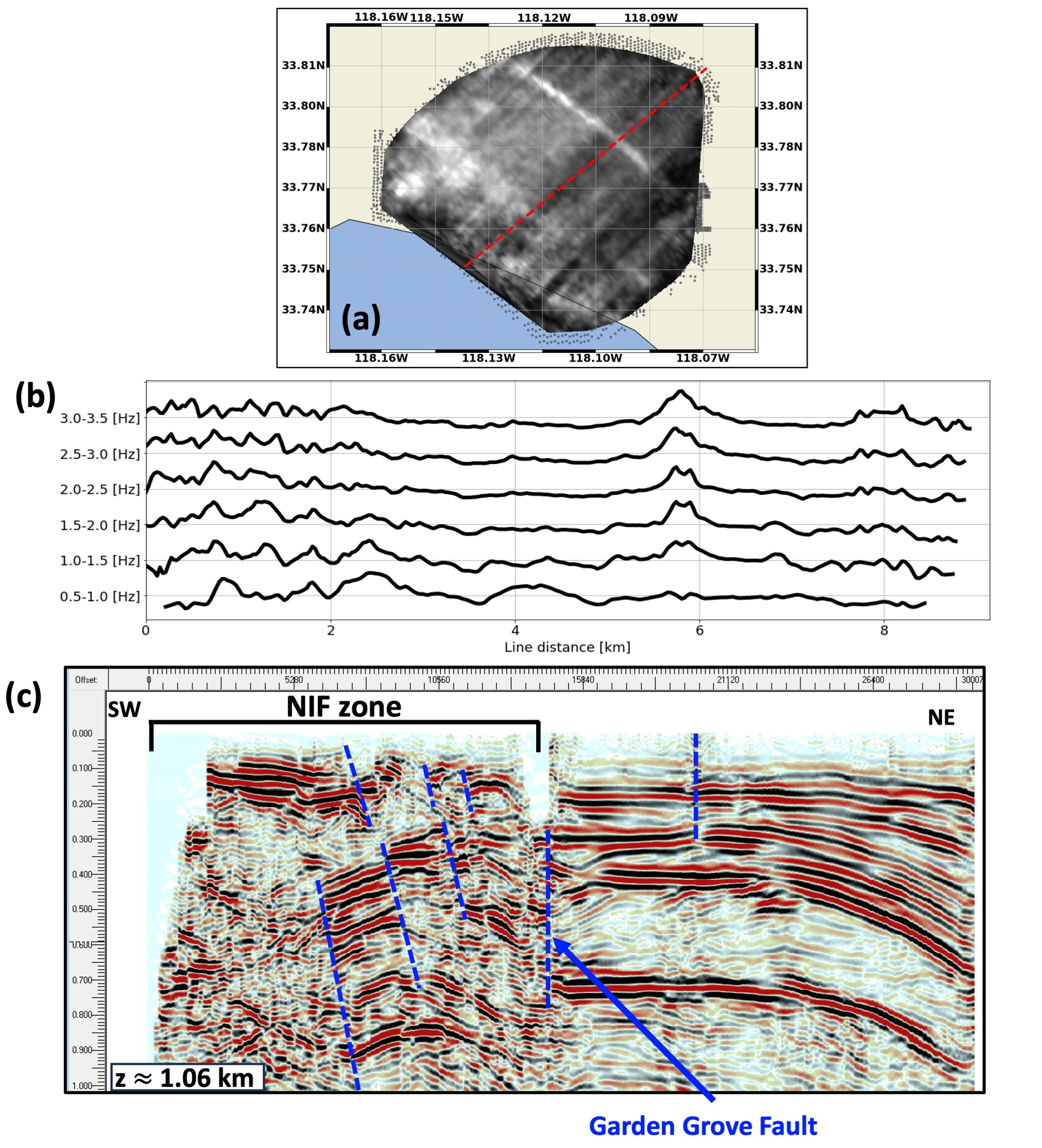}
\caption{Active survey profile and precursor image line at various frequencies. (a) Map view of the considered line for comparison (red dashed line). (b) Image line extracted from the precursory energy stacking process at various frequency bands. (c) Active-survey seismic image profile along the same line from panel (a)~\cite{gish2023sealbeach}. The blue dashed lines represent the interpreted faults.}
\label{fig:figure4}
\end{figure}

\begin{figure}[ht]
\centering
\includegraphics[width=0.9\linewidth]{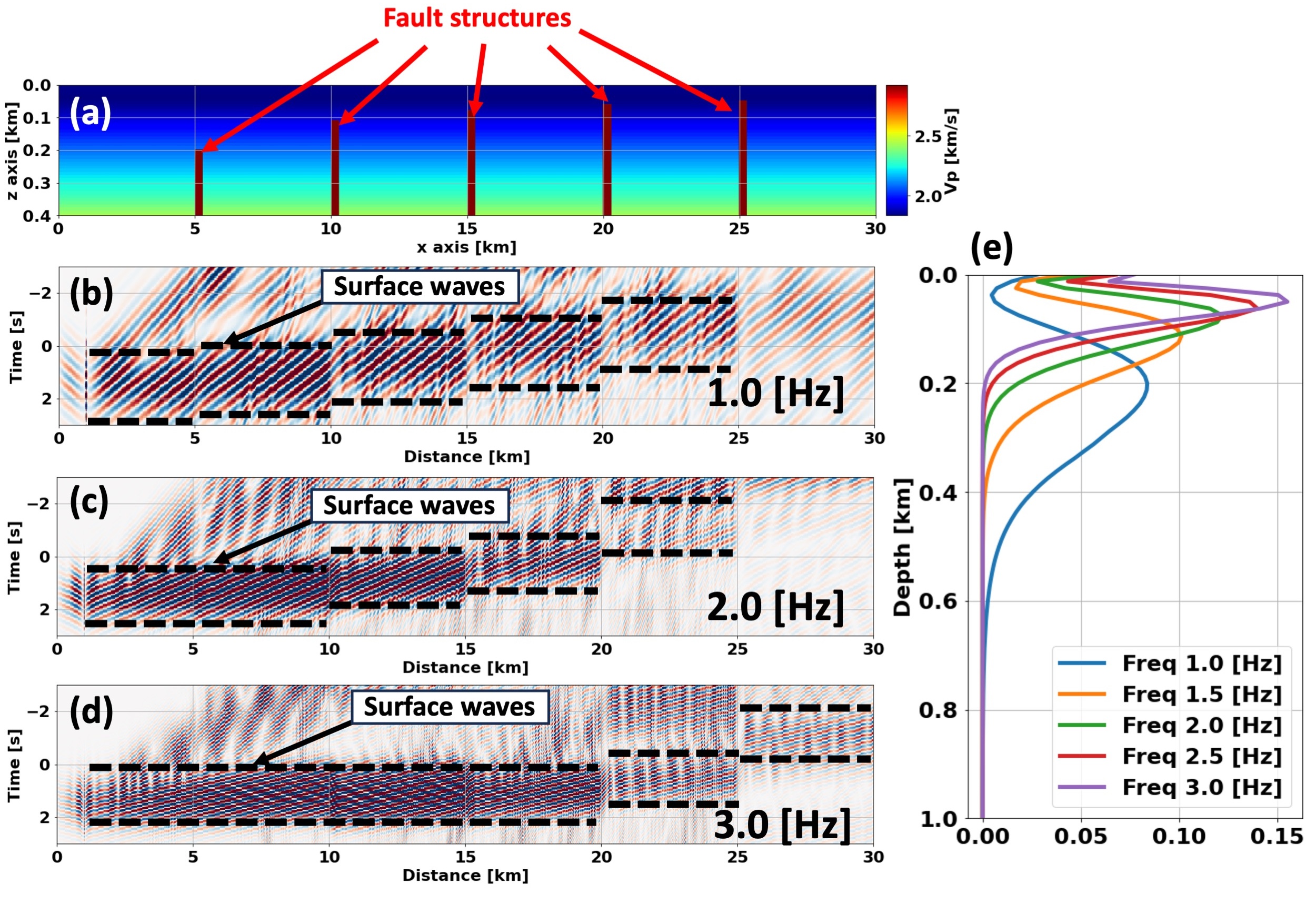}
\caption{Synthetic model test to highlight depth sensitivity from Rayleigh surface waves. (a) Synthetic velocity model used in the depth sensitivity test. (b-d) Vertical component data windows along the surface wave direct arrival filtered at different frequency bands. The dashed black lines denote the portion of the data containing the shifted surface waves. (e) Rayleigh surface-wave sensitivity kernels.}
\label{fig:figure5}
\end{figure}

\end{document}